\documentclass[titlepage,a4paper,11pt]{article}
\usepackage[utf8]{inputenc}
\usepackage{amssymb}
\usepackage{amsmath}

\usepackage{graphicx}
\usepackage{epstopdf}
\usepackage{fancyref}
\usepackage{hyperref}

\def\DD{{\cal D}}

\def\GG{{\cal G}}
\def\HH{{\cal H}}

\def\LL{{\cal L}}

\def\NN{{\cal N}}

\def\SS{{\cal S}}

\def\Tr{{\rm {Tr}}}

\def\d{{\partial}}

\def\beq{\begin{equation}}
\def\eeq{\end{equation}}
\newcommand{\bea}{\begin{eqnarray}}
\newcommand{\eea}{\end{eqnarray}}
\def\bal{\begin{align}}
\def\eal{\end{align}}

\begin{document}
\begin{titlepage} 
\begin{flushright}
CCQCN-2015-87\\
CCTP-2015-11
\vskip -1cm
\end{flushright}
\vskip 2.5cm
\begin{center}
    \centerline{\bf \LARGE Defects in Chern-Simons theory,}
      \vspace{0.4cm}
    \centerline{\bf \LARGE gauged WZW models on the brane,}
\vspace{0.4cm}
    \centerline{\bf \LARGE and level-rank duality}

\vskip 1cm
{Adi Armoni$^\dagger$ and Vasilis Niarchos$^\natural$}\\
\vskip 0.5cm
\medskip
{\it $^{\dagger}$Department of Physics, Swansea University \\
Singleton Park, Swansea, SA2 8PP, UK}\\
\smallskip
{\it $^\natural$Crete Center for Theoretical Physics and}\\
{\it Crete Center for Quantum Complexity and Nanotechnology}\\
{\it Department of Physics, University of Crete, 71303, Greece}\\
\medskip
{a.armoni@swansea.ac.uk, 
niarchos@physics.uoc.gr}

\end{center}
\vskip .3cm
\centerline{\bf Abstract}

\baselineskip 20pt
%

\vskip .5cm 
\noindent
We consider Hanany-Witten setups of 3- and 5-branes in type IIB string theory that realize 
$\NN=(1,0)$, $(2,0)$ and $(1,1)$ gauged WZW models in $1+1$ dimensions. The 
gauged WZW models arise as theories residing on the boundary of D3 branes ending on D5 branes.
From the point of view of low energy dynamics the D5 branes play the role of 
half-BPS co-dimension-1 defects (domain walls) in 3d $\NN=1$ or $\NN=2$ Chern-Simons theories. 
Extending the analysis of previous works on the subject of boundary conditions in (supersymmetric) 
Chern-Simons theory, we discuss in detail the field theory construction of a large class of Chern-Simons 
domain wall theories and its embedding in open string dynamics.
Finally, we exhibit how standard brane moves that result to 3d Seiberg
duality, translate in our setup to a generalized level-rank duality in gauged-WZW models.

\vfill
\noindent
\end{titlepage}\vfill\eject

\setcounter{equation}{0}

\pagestyle{empty}
\small
\vspace*{-0.7cm}
\tableofcontents
\normalsize
\newpage
\pagestyle{plain}
\setcounter{page}{1}

\section{Introduction} 
\label{setups}
 
We are interested in Hanany-Witten (HW) brane configurations \cite{Hanany:1996ie}  
that preserve one or two real supersymmetries and realize at low energies three-dimensional supersymmetric
Chern-Simons (CS) theories with a half-BPS co-dimension-1 defect (boundary or domain wall).
The CS theories reside on D3-branes suspended between two stacks of 5-branes and the 
co-dimension-1 defect lies at the intersection of the D3-branes with D5-branes.

Half-BPS domain walls and boundaries in 3d supersymmetric gauge 
theories are interesting from a quantum field theory point of view (see 
\cite{Gadde:2013wq,Okazaki:2013kaa,Sugishita:2013jca,Yoshida:2014ssa} for a sample of recent discussions), 
but also because of the information that they carry about D- and M-brane dynamics in string/M-theory
(an example of prominent interest involves M2-branes ending on M5-branes 
\cite{Strominger:1995ac,Townsend:1995af}). 

In the present work the 3d theories of interest are topological CS theories with $\NN=0,1,2$ supersymmetry. 
CS theories on spaces with boundary have been studied extensively in the past starting from the seminal work 
of Witten \cite{Witten:1988hf}. It is well known, in particular, that with suitable boundary conditions the 
boundary theory is a Wess-Zumino-Witten (WZW) model \cite{Elitzur:1989nr}. 
Our main contribution to this story can be summarized as follows:
\begin{itemize}
\item[$(a)$] We give a boundary degrees of freedom reformulation of the boundary conditions \cite{Moore:1989yh}
that realize gauged WZW models. In the process, we find it useful to formulate an extended class of domain wall 
theories in CS theory that has a natural embedding in brane setups.
\item[$(b)$] In $\NN=1,2$ supersymmetric CS theory we consider half-BPS domain wall theories that are given
by $\NN=(1,0)$, or $\NN=(1,1), (2,0)$ gauged WZW models. The boundary effects of supersymmetry are treated 
using the formalism of Ref.\ \cite{Belyaev:2008xk}, however, our approach differs from previous work on this subject 
in two ways. First, we treat gauge invariance differently from \cite{Berman:2009kj} by suitably incorporating 
boundary degrees of freedom along the lines of a generalization of \cite{Chu:2009ms,Faizal:2011cd}. 
Second, compared to \cite{Faizal:2011cd} we describe the case of 3d $\NN=2$ supersymmetry in $\NN=2$ 
superspace formalism without going to the so-called Ivanov gauge \cite{Ivanov:1991fn}.
\end{itemize}
The relevant constructions in field theory are discussed in sections \ref{bosonic}-\ref{N2}. 
Section \ref{bosonic} analyses a class of bosonic prototype cases. It captures the gist of the construction 
without going into the subtleties of supersymmetry.
Domain walls in $\NN=1$ CS theory are discussed in section \ref{N1}. The case of $\NN=2$ CS theory is
explained in section \ref{N2}.

The more interesting case of Chern-Simons-matter theories can be considered with similar methods by 
incorporating matter to our discussion. The relevant construction of boundary actions in the context of the ABJM 
theory \cite{Aharony:2008ug} and the orthogonal M2-M5 intersection will be discussed elsewhere.

As we mentioned already, suitable brane configurations in string theory can provide
an interesting perspective on the physics of domain wall theories in supersymmetric gauge theories.
Along these lines: 
\begin{itemize}
\item[$(a')$] We present brane setups that realize at low energies half-BPS domain walls in supersymmetric 
CS theories, and we argue that the corresponding domain wall theories are the ones analyzed in 
sections \ref{N1}, \ref{N2}. 
\item[$(b')$] A standard brane deformation that exchanges the position of 5-branes acts on 
the low-energy 3d theory as Seiberg duality \cite{Giveon:2008zn,Niarchos:2008jb}. 
We argue that the corresponding effect on the 
domain wall theories is a generalized level-rank duality.
\end{itemize}

The precise properties of open string dynamics in the brane configurations of interest will be considered in sections 
\ref{N1brane}, \ref{N2brane}. In the rest of this section we would like to give an up front summary of
the main configurations that will be discussed.

\section*{Brane setups}

We consider brane configurations in type IIB string theory that involve D3, D5, NS5 and 
$(1,k)$5-branes\footnote{We use conventions where
$(p,q)5$ denotes a fivebrane bound state with $p$ units of NS5-brane charge and $q$ units of D5-brane charge.
Without loss of generality we will henceforth assume that $k>0$.}
in ten-dimensional flat space. Different orientations of the D5 and the $(1,k)5$-brane bound state result in
different amounts of preserved supersymmetries. We will focus on the following three examples.

\subsection{1/32-BPS configuration: $d=2$, $\NN=(1,0)$}

The first example involves the brane orientations
\beq
\label{setupaa}
	\begin{array}{r c c c c c c c c c c c}
	N+M ~ D3_+ & ~:~ & 0  & 1 & 2_+ &  &  &  & |6| &  &  & 
	\\
	N ~ D3_- & ~:~ & 0  & 1 & 2_- &  &  &  & |6| &  &  & 
	\\
	1 ~ NS5  & ~:~ & 0  & 1 & 2 & 3 & 4 & 5 &  &  &  & 
	\\
	1 ~(1,k)5 & ~:~ & 0  & 1 & 2 & 3 &  & \left[ {5 \atop 9} \right]_{\theta} &  &  & 8 & 
	\\
	N+M ~ D5  & ~:~ & 0  & 1 &  & 3 & 4 & 5 & 6 &  &  & 
	\end{array}
\eeq
This table lists the number and type of branes involved and the directions along which they are extended. 
The notation $| 6 |$ denotes that a brane is oriented along the direction $6$ in a finite interval.
Here the D3-branes are suspended between the NS5-brane and the $(1,k)5$-brane which 
are separated along the direction 6. The notation $D3_\pm$ and $2_\pm$ 
denotes that the D3-branes are extended along the half-line $x^2>0$ for $+$, and $x^2<0$ for $-$.
Both sets of D3-branes end at $x^2=0$ at $N+M$ D5 branes. 
Finally, the notation $\left[ {5 \atop 9} \right]_\theta$ denotes that  
a brane stretches in the (59)-plane along a line at angle $\theta$ from the 5-axis. 
The quoted supersymmetry is preserved when the 
angle $\theta$ is fixed in terms of $k$ and the string coupling constant $g_s$
via the relation $\tan\theta = k\, g_s$.

In appendix \ref{braneSUSY} we show that 
the above configuration preserves only one real supersymmetry realizing a 2d $\NN=(1,0)$ theory at
the intersection along the (01)-plane. Away from the overlapping $N+M$ D5-branes the low energy theory at
the $(2+1)$-dimensional intersection is pure $\NN=1$ CS theory \cite{Kitao:1998mf}. 
The D5 intersection is a co-dimension-1 defect from the point of view of the CS theory. 
Across the defect the rank of the gauge group of the CS theory changes. The common direction 3 of the 
5-branes gives a classically massless $\NN=1$ scalar superfield in the 3d bulk theory, but as was pointed out in \cite{Armoni:2005sp} quantum effects make the non-abelian part of this multiplet massive and irrelevant for the 
deep infrared (IR) physics.

\subsection{1/16-BPS configuration: $d=2$, $\NN=(2,0)$}
\label{20setup}

A slightly different configuration that preserves two real supersymmetries and realizes a 2d $\NN=(2,0)$ theory has
the following ingredients
\beq
\label{setupab}
	\begin{array}{r c c c c c c c c c c c}
	N+M ~ D3_+ & ~:~ & 0  & 1 & 2_+ &  &  &  & |6| &  &  & 
	\\
	N ~ D3_- & ~:~ & 0  & 1 & 2_- &  &  &  & |6| &  &  & 
	\\
	1 ~ NS5  & ~:~ & 0  & 1 & 2 & 3 & 4 & 5 &  &  &  & 
	\\
	1 ~(1,k)5 & ~:~ & 0  & 1 & 2 & \left[ {3\atop 7}\right]_{-\theta} &  &  &  &  & 8 & 9
	\\
	N+M ~ D5  & ~:~ & 0  & 1 &  & 3 & 4 & 5 & 6 &  &  & 
	\end{array}
\eeq
In this case, the low energy 3d bulk theory away from the D5-brane stack along the (012)-plane is $\NN=2$ CS
theory. The theory at the co-dimension-1 defect at $x^2=0$ preserves $\NN=(2,0)$ supersymmetry.

\subsection{1/16-BPS configuration: $d=2$, $\NN=(1,1)$}
\label{11setup}

Finally, by changing the orientation of the D5-branes we can obtain a non-chiral $d=2$ $\NN=(1,1)$
theory at the defect
\beq
\label{setupac}
	\begin{array}{r c c c c c c c c c c c}
	N+M ~ D3_+ & ~:~ & 0  & 1 & 2_+ &  &  &  & |6| &  &  & 
	\\
	N ~ D3_- & ~:~ & 0  & 1 & 2_- &  &  &  & |6| &  &  & 
	\\
	1 ~ NS5  & ~:~ & 0  & 1 & 2 & 3 & 4 & 5 &  &  &  & 
	\\
	1 ~(1,k)5 & ~:~ & 0  & 1 & 2 & \left[ {3\atop 7}\right]_{-\theta} &  &  &  &  & 8 & 9
	\\
	N+M ~ D5  & ~:~ & 0  & 1 &  &  &  &  & 6 & 7 & 8 & 9
	\end{array}
\eeq
The supersymmetries preserved by this configuration are verified in appendix \ref{braneSUSY}.

\vspace{0.3cm}
Our main interest in all of the above cases is the low-energy domain wall theory that describes the 
dynamics at the two-dimensional D3-D5 intersection at $x^2=0$. Before we analyze this theory it will first
be useful to revisit separately the issue of domain walls and boundaries in CS theory. We proceed to discuss
three relevant examples with increasing amounts of supersymmetry, $\NN=0,1,2$.

\section{Chern-Simons domain walls}
\label{bosonic}

Our prototype for the more involved supersymmetric theories that follow is standard bosonic
Chern-Simons theory in $2+1$ dimensions (parametrized by coordinates $(x^0,x^1,x^2)$) 
in a slightly uncommon situation where the gauge group jumps abruptly on a $(1+1)$-dimensional 
defect located, say, at $x^2=0$. To be specific, consider on the left ($x^2<0$) CS theory at level $k$ with 
gauge group $\HH$, and on the right $(x^2>0)$ CS theory at level $k$ with gauge group $\GG$. 
We assume $\HH \subset \GG$. The total bulk action of the system at $x^2 \neq 0$ is
\beq
\label{bosonaa}
\SS_{bulk}[A,B] = \frac{k}{4\pi} \int_{x^2<0} \omega_3 (B) 
+ \frac{k}{4\pi} \int_{x^2>0} \omega_3 (A)
\eeq 
where we define the 3-form
\beq
\label{bosonab}
\omega_3 (A) = \Tr \left( A dA+\frac{2}{3} A^3 \right)
~.
\eeq
$A$ is a gauge field in the adjoint of $\GG$ and $B$ is a gauge field in the adjoint of $\HH$.

With a parity transformation 
\beq
\label{bosonac}
\frac{k}{4\pi}\int_{x^2<0} \omega_3 (B) 
= - \frac{k}{4\pi}\int_{x^2>0} \omega_3 (B) 
~,
\eeq
hence in an equivalent form
\beq
\label{bosonad}
\SS_{bulk}[A,B] = \frac{k}{4\pi} \int_{x^2>0} \Big( \omega_3 (A) -  \omega_3 (B) \Big)
~.
\eeq 
In this form our theory has been reformulated as a boundary problem. 
In what follows, we will mostly work with the boundary formulation \eqref{bosonad}.

It is well known that the CS theory has a gauge anomaly on spaces with boundary \cite{Elitzur:1989nr}; 
under a gauge transformation the action changes by a non-vanishing surface contribution.
There are two standard ways to cancel this surface contribution. We can either impose suitable boundary conditions
on the gauge field, or add explicit new degrees of freedom at the boundary
that transform under a gauge transformation.

We proceed to describe a specific example that shows how boundary degrees of freedom deal with 
the gauge anomaly in the context of the bulk action \eqref{bosonad}.
Then, we demonstrate the equivalence of this construction with an alternative formulation 
in terms of a suitable set of boundary conditions.

\subsection{Boundary degrees of freedom}
\label{bosondof}

Following a slight generalization of the discussion in Ref.\ \cite{Chu:2009ms} we consider 
boundary degrees of freedom $g$ that are group elements in the larger gauge group 
$\GG$. For these degrees of freedom we postulate the boundary action 
\bal
\label{bosonba}
\SS_{bdy}[A,B,g] =& -\frac{k}{8\pi} \int_{x^2=0} d^2x
\left\{
\Tr \left[ \left( g^{-1} \DD_\mu^A g \right)^2 \right]
- \Tr \left[ \Pi_\HH \left( g^{-1} \DD_\mu ^B g \right)^2 \right]
\right\}
\nonumber\\
&+\frac{k}{4\pi} \int_{x^2\geq 0} 
\Big\{ 
\left[ \omega_3(A^g) - \omega_3(A) \right]
- 
\left[ \omega_3(\Pi_\HH (B^g)) - \omega_3(B) \right]
\Big\} 
~.
\end{align}
We use notation where $\DD_\mu^A$, $\DD_\mu^B$ are gauge covariant derivatives with respect to the 
bulk gauge fields $A$ and $B$ respectively, $\DD_\mu^A=\partial _\mu + A_\mu \, , \,\,\, \DD_\mu^B =\partial _\mu + B_\mu$. $\Pi_\HH$ is a projector of elements of the Lie algebra of group $\GG$ 
to elements of the Lie algebra of $\HH$, and 
\beq
\label{bosonbb}
A^g = g^{-1} A g+ g^{-1}dg
\eeq
is the $g$ gauge transformation of $A$ (similar expressions apply to $B$).

Both lines on the RHS of eq.\ \eqref{bosonba} are supported on the boundary $x^2=0$. 
In particular, the term on the second line, which arises from the subtraction of two three-dimensional contributions, 
is a total derivative \cite{Chu:2009ms}. In the definition of the second line the domain of the group elements $g$ 
is extended in the bulk, but since the final result has support only on the boundary this bulk extension is not unique. 
For instance, we can use different bulk extensions of $g$ for the gauge fields $A$ and $B$.

Notice that the first term on the RHS of eq.\ \eqref{bosonba}
is obviously gauge invariant. It introduces kinetic terms for the degrees of freedom $g$
that lead to a natural two-dimensional CFT on the boundary. Clearly, the boundary theory is not unique,
and these terms are part of the choice we are making.

With this construction it is obvious that the total bulk-boundary action 
\bal
\label{bosonbc}
\SS_{total}[A,B,g] =& -\frac{k}{8\pi} \int_{x^2=0} d^2 x
\left\{
\Tr \left[ \left( g^{-1} \DD_\mu^A g \right)^2 \right]
- \Tr \left[ \Pi_\HH \left( g^{-1} \DD_\mu ^B g \right)^2 \right]
\right\}
\nonumber\\
&+\frac{k}{4\pi} \int_{x^2\geq 0} 
\Big\{ 
 \omega_3(A^g) 
- 
 \omega_3(\Pi_\HH (B^g))
\Big\} 
\end{align}
is invariant under gauge transformations in the group $\GG$ of the form
\beq
\label{bosonbd}
A^g \to \left( A^g \right)^h = A^{hg}~, ~~ B^g \to \Pi_{\HH} \left[ \left( B^g \right)^h \right] = \Pi_\HH \left[ B^{hg} \right]
~, ~~ g \to h^{-1} g~, ~~ h,g \in \GG
~.
\eeq

The bulk action has originally a $\GG \times \HH$ gauge anomaly at the boundary, and 
the second line in the boundary action \eqref{bosonba} cancels the ${\GG \over \HH} \times \HH_{vector}$ part 
expressed by the transformation \eqref{bosonbd}. ${\GG\over \HH}$ refers to gauge transformations of the
gauge field $A$ in the complement of $\HH$ and $\HH_{vector}$ refers to the vector part of 
$\HH_A \times \HH_B$, where $\HH_A$ acts on $A$ and $\HH_B$ acts on $B$. The axial part of 
$\HH_A\times \HH_B$ remains broken at the boundary.

Before we move on, it is instructive to consider the more explicit form of the total action \eqref{bosonbc}. 
Introducing light-cone coordinates $x^\pm = x^0\pm x^1$ and expanding out the boundary part \eqref{bosonba} 
we find
\bal
\label{bosonbe}
\SS_{bdy} =&~ S_{WZW,\GG_k}[g] - S_{WZW,\HH_k} [\Pi_\HH(g)]
\nonumber\\
&+ \frac{k}{4\pi} \int_{x^2=0} d^2 x
\Big\{
\Tr \Big[ \Pi_\HH \left( \d_+ g \, g^{-1} \left( A_- - B_- \right) \right) \Big]
+ \Tr \Big[ (1-\Pi_\HH)\left( \d_+ g\, g^{-1} \, A_- \right) \Big]
\Big\}
\nonumber\\
&+\frac{k}{8\pi} \int_{x^2=0} d^2 x\,
\left \{ \Tr \Big[ A_- A_+ \Big] - \Tr \Big[  B_- B_+ \Big] \right\}
~.
\end{align}
$S_{WZW,\GG_k}$ denotes the standard action of the WZW model at level $k$ with group $\GG$
\beq
\label{bosonbea}
S_{WZW,\GG_k}[g] = -\frac{k}{8\pi} \int_{x^2=0} d^2 x\, \Tr \left[ \left( g^{-1} \d g \right)^2 \right] 
-\frac{k}{12\pi} \int_{x^2\geq 0}  d^3x\, \Tr \left[ \left( g^{-1}\d g \right)^3 \right]
~.
\eeq
On the other hand, the bulk part \eqref{bosonad} can be recast into the following form after integrating by parts
\bal
\label{bosonbf}
\SS_{bulk} &= \frac{k}{4\pi} \int_{x^2>0} d^2x\,
\Tr \left[ A_+ F_{2-} +\frac{1}{2} \left( A_2 \d_+ A_- - A_- \d_+ A_2 \right) \right]
\nonumber\\
&- \frac{k}{4\pi} \int_{x^2>0} d^2x\,
\Tr \left[ B_+ G_{2-} +\frac{1}{2} \left( B_2 \d_+ B_- - B_- \d_+ B_2 \right) \right]
\nonumber\\
& -\frac{k}{8\pi} \int_{x^2=0} d^2x\,
\Tr \Big[ A_-A_+ -  B_-B_+ \Big] 
~.
\end{align}
$F_{\mu\nu}$ and $G_{\mu\nu}$ are respectively the field strengths of the non-abelian gauge fields $A_\mu$ and
$B_{\mu}$.

When we add together \eqref{bosonbe} and \eqref{bosonbf} to obtain $\SS_{total}$ the boundary term
$\int \Tr [ A_-A_+ - B_-B_+ ]$ cancels out and the gauge field components $A_+$, $B_+$ appear as 
Lagrange multipliers. Integrating them out we obtain $F_{2-}=0$, $G_{2-}=0$ that we solve by setting
\beq
\label{bosonbg}
A_i = U^{-1} \d_i U~, ~~ B_i = V^{-1} \d_i V~, ~~ i= - ,2~, ~~ U\in \GG~, ~~ V\in \HH
~.
\eeq
Then, employing the Polyakov-Wiegmann identity, and setting \cite{Karabali:1989dk}
\beq
\label{bosonbi}
\Pi_\HH (Vg) = h^{-1} \tilde h ~, ~~ Ug = h^{-1} g \tilde h ~, ~~ h ~, ~\tilde h \in \HH
\eeq
we find that $\SS_{total}$ is the vector $\GG_k/\HH_k$ gauged WZW action
\beq
\label{bosonbj}
\SS_{total} = S_{WZW,\GG_k}[g] +
\frac{k}{4\pi} \int_{x^2=0} d^2x\,
\Tr \Big[
- a_+ g^{-1} a_- g + a_+ g^{-1} \d_- g - \d_+ g g^{-1} a_- +a_+ a_- \Big]
\eeq
with gauge fields
\beq
\label{bosonbk}
a_- = \d_- \tilde h \, \tilde h^{-1}~, ~~ a_+ = \d_+ h \, h^{-1}
~.
\eeq

\subsection{Boundary conditions}

For later purposes it will be useful to know if the same final result can be obtained by using appropriate boundary 
conditions. It is known \cite{Moore:1989yh} that the bulk CS action \eqref{bosonad} admits the following 
boundary conditions 
\bal
\label{bosonca}
&B_\pm = \Pi_\HH \left ( A_\pm \right )
\\
\label{bosoncaa}
&(1- \Pi_\HH) (A_+) =0
~.
\end{align}
These conditions set the boundary term $\int \Tr [ A_-A_+ - B_-B_+ ]$ in \eqref{bosonbf} to zero and then by
standard manipulations analogous to the ones performed in the previous subsection they lead to the 
vector $\GG_k/\HH_k$ gauged WZW action \eqref{bosonbj}. We conclude that the boundary conditions 
\eqref{bosonca}, \eqref{bosoncaa} are equivalent to the boundary action \eqref{bosonba}. 
Notice in particular that, as in the case of \eqref{bosonca}, \eqref{bosoncaa}, the axial part of $\HH_A \times \HH_B$ 
is broken explicitly at the boundary by \eqref{bosonca}, but the vector part is preserved.

In the following sections we will choose to formulate domain walls and boundaries in supersymmetric CS theories 
using the approach of boundary degrees of freedom and boundary interactions. 
This approach provides a flexible uniform prescription for many cases that is convenient
for the resolution of issues related to supersymmetry and gauge invariance.

\paragraph{Note.} The Euler-Lagrange variation of the action imposes additional on-shell boundary conditions.
We will not discuss these conditions explicitly here. The relevant details can be found for example in Ref.\ 
\cite{Chu:2009ms}.

\subsection{An extended class of boundary actions}
\label{general}

The domain wall theory \eqref{bosonba} is by no means unique. The basic building block of \eqref{bosonba}
is the boundary action \cite{Chu:2009ms}
\beq
\label{generalaa}
\SS_{bdy}[A,g] = - \frac{k}{8\pi} \int_{x^2=0} d^2x \, \Tr \left[ \left(g^{-1} \DD_\mu g \right)^2 \right]
+\frac{k}{4\pi} \int_{x^2\geq 0} \Big [  \omega_3(A^g) - \omega_3(A) \Big ]
\eeq
written here for a single bulk gauge field $A$ in the Lie algebra of a group $\GG$, and $g$ an element of 
the same group. In section \ref{N1brane} we will encounter a generalization of this construction that arises 
naturally from open string dynamics. It will be useful to describe this extension here in a simplified 
non-supersymmetric, bosonic context. 

The crucial feature that makes \eqref{generalaa} work is the fact that $g$ transforms as a field in the 
fundamental representation of the bulk gauge group (under the left action of the group). As is evident
from \eqref{bosonbd} the simultaneous left action of the group on $g$ with a bulk gauge transformation is
enough to render the combination $A^g$ invariant. The passive role of the right action of the gauge group 
in this manipulation suggests a natural generalization, where instead of considering boundary degrees of 
freedom $g$ in the bi-fundamental of 
$\GG_L \times \GG_R$,\footnote{$\GG_L$ $(\GG_R)$ represents the action of $\GG$ from the left (right).}
we consider them in the bi-fundamental of the general product $\GG_L \times \GG'$.
$\GG'$ can be different from $\GG$. Accordingly, in this more general case, we will denote the Hermitian 
conjugate of $g$ by $\bar g$, instead of $g^{-1}$. 

With these specifications, we can construct an extended class of boundary actions
\beq
\label{generalab}
\SS_{bdy}[A,g] = - \frac{k}{8\pi} \int_{x^2=0} d^2x \, \Tr \left[ \left(\bar g\, \DD_\mu g \right)^2 \right]
+\frac{k}{4\pi} \int_{x^2> 0} \left [  \omega_3({\boldsymbol A}^g) - \omega_3(A) \right]
\eeq
where ${\boldsymbol A}^g$ in the Lie algebra of $\GG'$ denotes the combination
\beq
\label{generalac}
{\boldsymbol A}^g \equiv \tilde g\, A \, g + \tilde g\,  d g
~.
\eeq
$\tilde g$ is defined as
\beq
\label{generalaca}
\tilde g = \bar g \left( g \bar g \right)^{-1}
\eeq
with the property $g \tilde g = {\bf 1}_{\GG'}$.
We are using bold fonts for ${\boldsymbol A}^g$ to distinguish it from the standard bulk gauge transformation 
$A^h$, $h\in \GG$.

With these definitions, \eqref{generalab} continues to exhibit the nice features of \eqref{generalaa}. It is a 
boundary action for $g$ independent of its bulk extension. Moreover,
${\boldsymbol A}^g$, as well as the total bulk-boundary action
\beq
\label{generalaba}
\SS_{total}[A,g] = - \frac{k}{8\pi} \int_{x^2=0} d^2x \, \Tr \left[ \left(\bar g\, \DD_\mu g \right)^2 \right]
+\frac{k}{4\pi} \int_{x^2> 0} \omega_3({\boldsymbol A}^g)
~,
\eeq
are invariant under the combined bulk-boundary gauge transformation
\beq
\label{generalad}
A \to A^h ~, ~~ g \to h^{-1} g~, ~~ \bar g \to \bar g \, h~, ~~ h\in \GG
\eeq
as desired.

In section \ref{N1brane} the boundary action \eqref{bosonba} will 
be recovered from this more general construction as a special case where extra massive
boundary degrees of freedom are integrated out to produce naturally the terms that involve the projection on the 
subgroup $\HH$.

\section{$\NN=1$ Chern-Simons theory}
\label{N1}

We proceed to describe the $\NN=1$ supersymmetric version of the previous discussion. Compared to the 
bosonic case, where we had to worry only about the gauge symmetry, here we also have to consider
what happens to the supersymmetry. In general, the co-dimension-1 defect breaks the bulk supersymmetry,
and since we are interested in half-BPS defects some additional care needs to be taken to ensure that the 
appropriate amount of supersymmetry is restored on the defect by suitable boundary interactions.

\subsection{Details of $\NN=1$ supersymmetry}

It is convenient to work in 3d $\NN=1$ superspace formalism with coordinates $(x^\mu,\theta_\alpha)$. 
Our conventions are summarized in appendix \ref{SUSYconventions}. 

The supersymmetric multiplet that contains the gauge field can be packaged in 
a spinor superfield $\Gamma_\alpha$ that contains a Majorana spinor $\chi_\alpha$, a real scalar $M$,
the gaugino $\lambda_\alpha$ and the gauge field $A_\mu$
\beq
\label{N1aa}
\Gamma_\alpha = \chi_\alpha + \theta_\alpha M+ (\gamma^\mu \theta)_\alpha A_\mu 
+\theta^2 \left( \lambda_\alpha - \left( \gamma^\mu \d_\mu \chi\right)_\alpha \right)
~.
\eeq
The $\NN=1$ CS theory at level $k$ with gauge group $\GG$ takes the form
\bal
\label{N1ab}
\SS_{\GG_k}[\Gamma] &=  \frac{k}{4\pi} \int d^3 x \, d^2\theta \,
\Tr \left[ \Gamma^\alpha \Omega_\alpha \right] 
\nonumber\\
& = \frac{k}{4\pi} \int d^3 x\, \Tr \Big[
\varepsilon^{\mu\nu\rho} \left( A_\mu \d_\nu A_\rho +\frac{2}{3} A_\mu A_\nu A_\rho \right)
+ \lambda^\alpha \lambda_\alpha + \DD_\mu \left( \chi^\alpha \left( \gamma^\mu \right)_\alpha^{~\beta} 
\lambda_\beta \right)
\Big]
~.
\end{align}
Specific expressions for the spinor superfield $\Omega_\alpha$ in terms of $\Gamma_\alpha$ are provided in 
appendix \ref{SUSYconventions}.

\paragraph{Supersymmetry restoring boundary interactions.}
We will introduce boundary degrees of freedom and boundary interactions that restore 
half of the bulk supersymmetry along the lines of \cite{Belyaev:2008xk,Berman:2009kj}.
More specifically, it has been shown \cite{Belyaev:2008xk} that the general bulk-boundary action 
\beq
\label{N1ac}
\int d^3 x \, \Big( d^2 \theta \, \LL \pm \d_2 \LL \big |_{\theta=0} \Big)
\eeq
preserves the supersymmetry generated by the supercharge 
$Q_\mp$. In what follows, we choose, by convention,
to preserve the supersymmetry generated by $Q_-$.
The domain wall theory is a 2d $\NN=(1,0)$ theory.

Applying the prescription \eqref{N1ac} to the $\NN=1$ CS theory \eqref{N1ab} we obtain the action
\bal
\SS_{{\tt SUSY},\GG_k}[\Gamma] &= 
\frac{k}{4\pi} \int d^3 x \, \Big( d^2\theta ~ \Tr \left[ \Gamma^\alpha \Omega_\alpha \right] 
+ \d_2 \left( \Gamma^\alpha \Omega_\alpha \right) \big |_{\theta=0} \Big) 
\nonumber\\
&= \frac{k}{4\pi} \int d^3 x\, \Tr \Big[
\varepsilon^{\mu\nu\rho} \left( A_\mu \d_\nu A_\rho +\frac{2}{3} A_\mu A_\nu A_\rho \right)
+ \lambda^\alpha \lambda_\alpha + \DD_\mu \left( \chi^\alpha \left( \gamma^\mu \right)_\alpha^{~\beta} 
\lambda_\beta \right)
\Big]
\nonumber\\
&~ - \frac{k}{4\pi} \int d^3 x\, \d_2 \Tr 
\Big[ 
\chi^\alpha  \lambda_\alpha + \frac{1}{2} \chi^\alpha 
\left[ \left( \gamma^\mu A_\mu\right)_\alpha^{~\beta},\chi_\beta \right] 
\Big]
~.
\end{align}
We notice that the auxiliary field $M$ (see eq.\ \eqref{N1aa}) is absent from this action.

\subsection{$\NN=(1,0)$ gauged WZW models}

We are now in position to formulate the $\NN=1$ supersymmetric version of subsection \ref{bosondof}.
In the bulk $(x^2>0)$ we have $\NN=1$ CS theory with gauge group $\GG$ at level $k$ and 
$\NN=1$ CS theory with gauge group $\HH\subset \GG$ at level $-k$. We will denote the corresponding
spinor superfields $\Gamma_\alpha^\GG$ and $\Gamma_\alpha^\HH$. With the inclusion of the 
supersymmetrizing boundary interactions we denote
\beq
\label{N1ba}
\SS_{bulk} [\Gamma^\GG,\Gamma^\HH ] 
= \SS_{{\tt SUSY},\GG_k} [\Gamma^\GG] + \SS_{{\tt SUSY},\HH_{-k}} [\Gamma^\HH]
~.
\eeq
$\SS_{bulk}$ is invariant under $Q_-$, but gauge symmetry is broken at the boundary.

In complete analogy to the bosonic case of subsection \ref{bosondof}
we restore the ${\GG \over \HH} \times \HH_{vector}$ part of the gauge symmetry by introducing 
boundary $\NN=1$ scalar multiplets $g$ valued in the larger gauge group $\GG$ 
\beq
\label{N1bb}
g(x,\theta) = {\boldsymbol g}(x) +\theta^\alpha {\boldsymbol \psi}_\alpha(x) + \theta^2 {\boldsymbol f}(x)
\eeq
with the boundary action
\bal
\label{N1bc}
\SS_{bdy}\left[ \Gamma^\GG, \Gamma^\HH , g \right] 
&= \SS_{kin,WZW_k} \left[ g, \Gamma^\GG \right] 
+ \SS_{{\tt SUSY},\GG_k} [\left( \Gamma^\GG \right)^g ] - \SS_{{\tt SUSY},\GG_k} [\Gamma^\GG]
\nonumber\\
&~+ \SS_{kin,WZW_{-k}} \left[ \Pi_\HH(g), \Gamma^\HH \right] 
+ \SS_{{\tt SUSY},\HH_{-k}} \left [ \Pi_\HH \left( \left( \Gamma^\HH \right)^g \right) \right ] 
- \SS_{{\tt SUSY},\HH_{-k}} [\Gamma^\HH]
~.
\end{align}
$\Gamma^g$ denotes the gauge transformed spinor multiplet
\beq
\label{N1bd}
\Gamma^g_\alpha = g \nabla_\alpha g^{-1}
~,
\eeq
where $\nabla_\alpha$ is a super-gauge-covariant derivative defined in appendix \ref{SUSYconventions}.
For the component fields $(\chi_\alpha, A_\mu, \lambda_\alpha)$ of each of the 
spinor multiplets $\Gamma^\GG$ and $\Gamma^\HH$ that appear in \eqref{N1bc} 
the gauge transformation \eqref{N1bd} acts as follows
\bal
\label{N1be}
(\chi_\alpha)^g &= {\boldsymbol g} \, \chi_\alpha \, {\boldsymbol g}^{-1} - {\boldsymbol \psi}_\alpha \, 
{\boldsymbol g}^{-1}~,
\\
(A_\mu)^g &= {\boldsymbol g}\, A_\mu \, {\boldsymbol g}^{-1} + {\boldsymbol g}\, \d_\mu {\boldsymbol g}^{-1}~,
\\
(\lambda_\alpha)^g &= {\boldsymbol g} \, \lambda_\alpha \, {\boldsymbol g}^{-1}
~.
\end{align}
Finally, in \eqref{N1bc} $\SS_{kin,WZW_k}[g,\Gamma]$ denotes the gauge-covariant kinetic term that appears in the 
$\NN=(1,0)$ WZW model \cite{Howe:1987qv,Hull:1990qf}. Denoting by 
\beq
\label{N1bf}
\hat g = g(x,\theta)\big |_{x^2=0,\, \theta_+ =0} 
\eeq
the boundary $\NN=(1,0)$ projection of the group superfield $g$, we can write
\beq
\label{N1be}
\SS_{kin, WZW_k}[g,\Gamma] =  
-\frac{k}{2\pi} \int d^2 x \int d\theta_- \left ( \hat g^{-1} \hat \nabla_- \hat g \right) \left(\hat g^{-1} \DD_+ \hat g \right)
~.
\eeq
$\hat \nabla_\alpha$ denotes the boundary version of the super-gauge-covariant derivative, where all contributions
are evaluated at the boundary and projected on the appropriate chirality \cite{Faizal:2011cd}. 

Putting these formulae together we obtain the $\NN=1$ supersymmetric version of subsection \ref{bosondof}.
The total action 
\beq
\label{N1bf}
\SS_{total}\left[ \Gamma^\GG, \Gamma^\HH, g \right] = \SS_{bulk}\left[ \Gamma^\GG, \Gamma^\HH \right] 
+\SS_{bdy} \left[ \Gamma^\GG, \Gamma^\HH, g \right]
\eeq
provides the $\NN=(1,0)$ supersymmetric completion of the vector $\GG_k / \HH_k$ gauged WZW model.


\section{$\NN=2$ Chern-Simons theory}
\label{N2}

The extension to $\NN=2$ CS theory can be performed in a similar fashion. With $\NN=2$ supersymmetry
there are two types of half-BPS domain walls: the first type preserves $\NN=(2,0)$ supersymmetry on the
defect and the second type $\NN=(1,1)$ supersymmetry.

\subsection{Details of $\NN=2$ supersymmetry}

Our conventions for $\NN=2$ supersymmetry are summarized in appendix \ref{SUSYconventions}.
Before we delve into the details of our construction it will be useful to highlight a few well known details of $\NN=2$
supersymmetry that one should be aware of.
 
The $\NN=2$ superspace has two sets of $\NN=1$ Grassmann variables, 
($\theta_{1\alpha}$, $\theta_{2\alpha}$), that can be combined into the complex Grassmann variables
$$
\vartheta_\alpha = \frac{1}{\sqrt 2} \left( \theta_{1\alpha} + i \, \theta_{2\alpha} \right)
~.
$$
Accordingly, the component fields of an $\NN=2$ vector multiplet can be arranged in an $\NN=2$ superfield $V$
that can be built out of two $\NN=1$ scalar superfields $A$, $B$ and an $\NN=1$ spinor superfield 
$\Gamma_\alpha$ as
\beq
\label{N2aa}
V(x, \theta_1,\theta_2) = A(x,\theta_1) + \theta_2^\alpha \Gamma_{\alpha}(x,\theta_1) 
+ \theta_2^2 \left( B(x,\theta_1) + D_1^2 A (x,\theta_1) \right)
~.
\eeq

The $\NN=2$ CS theory at level $k$ with gauge group $\GG$ has a four-dimensional 
formulation of the form \cite{Ivanov:1991fn}
\beq
\label{N2ab}
\SS_{\GG_k} [V] = -
\frac{k}{2\pi} \int d^3 x \int d^4 \vartheta \int_0^1 ds \, 
\Tr \left[ V \bar {\boldsymbol D}^\alpha \left( e^{sV} {\boldsymbol D}_\alpha \, e^{-sV} \right) \right]
~.
\eeq
${\boldsymbol D}_\alpha$ is the $\NN=2$ superspace covariant derivative \eqref{2conaba}.

$\NN=2$ gauge transformations
\beq
\label{N2ac}
V \to V^g~, ~~{\rm such~that}~~ e^{V^g} = \bar g\,  e^V \, g
~,
\eeq
with $g$ an $\NN=2$ chiral superfield valued in the group $\GG$,
can be used to shift away the $\NN=1$ superfield $A$ in \eqref{N2aa} and to go to a convenient 
gauge where $A=0$ and the $\NN=2$ CS action \eqref{N2ab} simplifies to 
\beq
\label{N2ad} 
\SS_{\GG_k}[V]= \frac{k}{4\pi} \int d^3 x \int d^2\theta_1 \,  \Tr \left[ B^2 + \Gamma^\alpha \Omega_\alpha \right]
~.
\eeq
In this gauge (the so-called Ivanov gauge) the $B$ superfield is auxiliary and the 
$\NN=2$ CS action is essentially identical to the $\NN=1$ CS action. 

For an abelian gauge group $\GG$ the integral over $s$ in \eqref{N2ab} can be performed explicitly 
in the general gauge and the $\NN=2$ CS action can be expressed simply in three dimensions in terms 
of the $\NN=1$ superfields $A,B,\Gamma_\alpha$ \cite{Ivanov:1991fn}. 
For non-abelian $\GG$, however, there is no such simple expression and it is common to work instead in the 
Ivanov-Wess-Zumino gauge where $A=0$ and $\chi_\alpha=M=0$ in $\Gamma_\alpha$.

In our case, the presence of the defect obstructs this passage to the Ivanov gauge, unless we choose
to break explicitly part of the $\NN=2$ super-gauge invariance. In what follows we will opt to keep the full 
$\NN=2$ symmetry. That means we will have to work with the complete four-dimensional actions \eqref{N2ab}
without implementing the Ivanov gauge.

\paragraph{Supersymmetry restoring boundary interactions.}
The prescription of Ref.\ \cite{Belyaev:2008xk} can be applied to the general $\NN=2$ action
(see also \cite{Berman:2009kj})
\beq
\label{N2ae}
\SS_0 = \int d^3x \, d^2 \theta_1 \, d^2 \theta_2 \, \LL(\theta_1,\theta_2)
\eeq
to restore either $\NN=(2,0)$ or $\NN=(1,1)$ supersymmetry on the two-dimensional boundary. 
It is not hard to verify that the action
\beq
\label{N2af}
\SS^{(\pm)} = \int d^3x\, \Big\{ d^2\theta_1 \, d^2 \theta_2 \, \LL -  d^2\theta_1 \, \d_2 \LL \big |_{\theta_2=0} 
\pm \Big( - d^2\theta_2 \, \d_2 \LL \big |_{\theta_1=0} + \d_2 \d_2 \LL \big |_{\theta_1=\theta_2=0} \Big)
\Big\}
\eeq
preserves the supersymmetries generated by $\left(Q_{1\mp}, Q_{2-} \right)$. We will use the notation
$\SS^{(\pm)}_{{\tt SUSY},\GG_k} [V]$ to denote $\SS^{(\pm)}$ in the case of $\NN=2$ CS theory.

\subsection{$\NN=(2,0)$ gauged WZW models}

In the bulk $(x^2>0)$ we consider $\NN=2$ CS theory with gauge group $\GG$ at level $k$ and 
$\NN=2$ CS theory with gauge group $\HH$ at level $-k$. We will denote the corresponding 
$\NN=2$ vector multiplets $V^\GG$, $V^\HH$. Including the boundary interactions that restore the 
$(Q_{1-}, Q_{2-})$ supersymmetries we set
\beq
\label{N2ba}
\SS_{bulk}[V^\GG, V^\HH] = \SS^{(+)}_{{\tt SUSY},\GG_k}[V^\GG] 
+ \SS^{(+)}_{{\tt SUSY},\HH_{-k}}[V^\HH]
~.
\eeq

We restore the gauge symmetry that leads to the 2d $\NN=(2,0)$ $\GG_k /\HH_k$ coset on the boundary 
by introducing $\NN=2$ chiral multiplets $g$ valued in the larger gauge group $\GG$ with boundary action 
\bal
\label{N2bb}
\SS_{bdy} &=
\SS^{(2,0)}_{kin,WZW_k}\left[ g,V^\GG \right]
+\SS^{(+)}_{{\tt SUSY},\GG_k} \left[ (V^\GG)^g \right] - \SS^{(+)}_{{\tt SUSY},\GG_k} \left[ V^\GG \right] 
\nonumber\\
&~~~ +  \SS^{(2,0)}_{kin,WZW_{-k}}\left[ \Pi_\HH(g),V^\HH \right]
+\SS^{(+)}_{{\tt SUSY},\HH_{-k}} \left[ \Pi_\HH \left( (V^\GG)^g \right) \right] 
- \SS^{(+)}_{{\tt SUSY},\HH_{-k}} \left[ V^\HH \right] 
~.
\end{align}
The gauge covariantized $\NN=(2,0)$ 
WZW kinetic terms $\SS^{(2,0)}_{kin,WZW_k}[g,V]$ have the following form \cite{Howe:1987qv,Hull:1990qf}.
On the boundary we project the bulk group $\NN=2$ superfield $g$ into
\beq
\label{N2bc}
\hat g = g(x,\theta_1,\theta_2) \big |_{x^2=0, ~ \theta_{1+}=0,~\theta_{2+}=0}
\eeq
and define
\beq
\label{N2bd}
\SS^{(2,0)}_{kin,WZW_k}[g,V] = -\frac{k}{2\pi} \int d^2 x \int d\vartheta_- \, 
\left( \hat g^{-1} \hat {\boldsymbol \nabla}_- \hat g\right) \left( \hat g^{-1} \DD_+ \hat g \right)
~.
\eeq
$\hat {\boldsymbol \nabla}_\alpha$ is now the boundary version of the chiral $\NN=2$ super-gauge-covariant 
derivative (see appendix \ref{2conventions}).

\subsection{$\NN=(1,1)$ gauged WZW models}

The case of $\NN=(1,1)$ supersymmetry can be formulated in a similar fashion. The $(+)$ expressions are 
replaced by the $(-)$ expression and the $\NN=(2,0)$ WZW kinetic terms by $\NN=(1,1)$ WZW kinetic terms.
For quick reference we summarize the pertinent formulae:
\beq
\label{N2ca}
\SS_{bulk}[V^\GG, V^\HH] = \SS^{(-)}_{{\tt SUSY},\GG_k}[V^\GG] 
+ \SS^{(-)}_{{\tt SUSY},\HH_{-k}}[V^\HH]
~,
\eeq
\bal
\label{N2cb}
\SS_{bdy} &=
\SS^{(1,1)}_{kin,WZW_k}\left[ g,V^\GG \right]
+\SS^{(-)}_{{\tt SUSY},\GG_k} \left[ (V^\GG)^g \right] - \SS^{(-)}_{{\tt SUSY},\GG_k} \left[ V^\GG \right] 
\nonumber\\
&~~~ +  \SS^{(1,1)}_{kin,WZW_{-k}}\left[ \Pi_\HH(g),V^\HH \right]
+\SS^{(-)}_{{\tt SUSY},\HH_{-k}} \left[ \Pi_\HH \left( (V^\GG)^g \right) \right] 
- \SS^{(-)}_{{\tt SUSY},\HH_{-k}} \left[ V^\HH \right] 
~,
\end{align}
\beq
\label{N2cc}
\SS^{(1,1)}_{kin,WZW_k}[g,V] = -\frac{k}{2\pi} \int d^2 x \int d\theta_{1+} \, d\theta_{2-} \, 
\left( \hat g^{-1} \hat {\boldsymbol \nabla}_- \hat g\right) \left( \hat g^{-1} \hat{\boldsymbol \nabla}_+ \hat g \right)
~,
\eeq
with the boundary projection
\beq
\label{N2cd}
\hat g =  g(x,\theta_1,\theta_2) \big |_{x^2=0, ~ \theta_{1-}=0,~\theta_{2+}=0}
~.
\eeq

\section{1/32-BPS brane setups and level-rank duality}
\label{N1brane}

After the long technical introduction, we are finally in position to
discuss more explicitly the properties of the brane configurations \eqref{setupaa}.
Our main goal is to identify the low-energy gauge theory that resides on the semi-infinite D3-branes 
and to formulate the domain wall theory on D3-D5 intersections.

\subsection{Identification of the low-energy theory}
\label{IRtheory}

Brane configurations with D3-branes suspended along the direction 6
between NS5 and $(1,k)5$-branes, 
without the D5-branes across $(013456)$, 
have been studied extensively in the past \cite{Aharony:1997ju,Kitao:1998mf,Bergman:1999na}. 
The low-energy 3d theory on $N$ suspended D3-branes is $\NN=1$ CS theory at level $k$ 
with gauge group $U(N)$ coupled to a classically massless $\NN=1$ scalar multiplet in the adjoint 
representation that captures the classically free motion of the D3 branes in the common fivebrane
direction 3. As was emphasized in Ref.\ \cite{Armoni:2005sp} the traceless part of this multiplet 
receives a mass quantum mechanically, which is $1/N$ suppressed compared to the CS 
mass of the gauge field \cite{Armoni:2009vv}. With the assumption of a hierarchy between these 
two scales one can integrate out the adjoint scalar multiplet to obtain pure 
$\NN=1$ CS theory in the deep IR. The trace part of the adjoint scalar multiplet decouples as a 
free field. Accordingly, the $N$ suspended D3-branes in this setup behave as a bound state that can move
freely along the direction 3.

\begin{figure}[t!]
   \centering
   \includegraphics{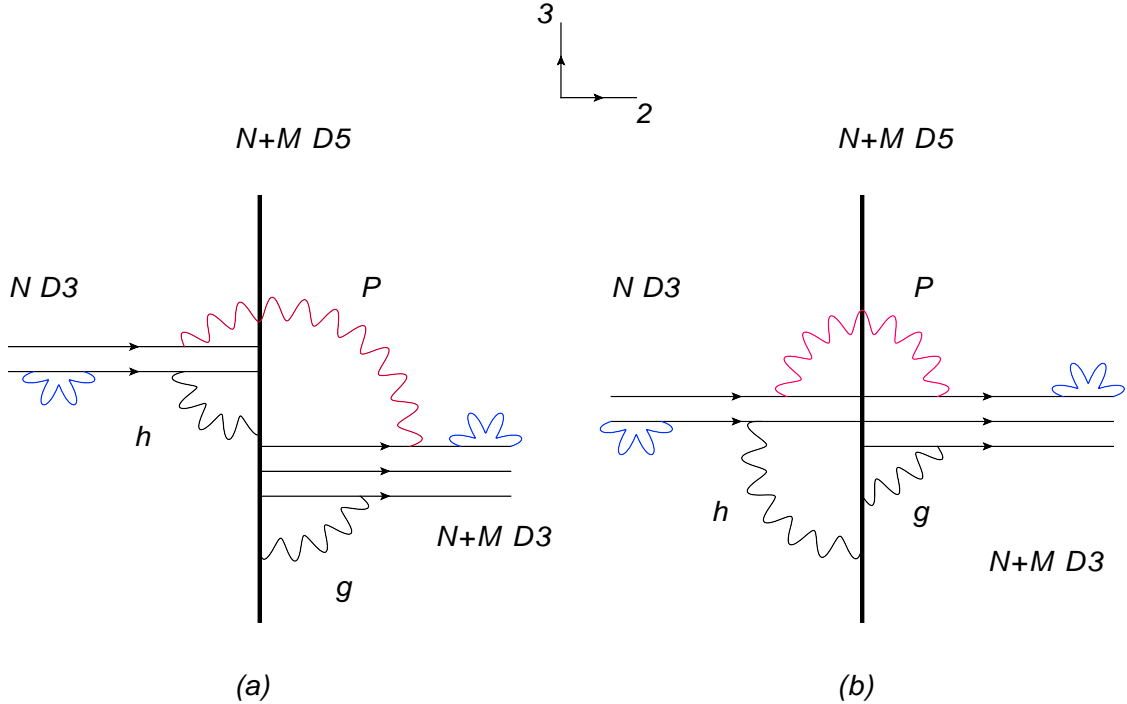}
    {\bf \caption{\rm Two separate stacks of D3 branes end from the left $(x^2<0)$ and the right $(x^2>0)$ on 
    D5-branes within the brane configuration \eqref{setupaa}. In Fig.\ $(a)$ the two D3 stacks are located 
    in different positions in the 3-direction, whereas in Fig.\  $(b)$ they are located at the same $x^3$ position.
    }
   \label{D3D5}
   }
\end{figure}

The $N+M$ D5-branes along $(013456)$ introduce new features. 
Two different sets of D3-branes can end on the D5-branes from the left ($x^2<0$) and the right ($x^2>0$)
as summarized in Fig.\ \ref{D3D5}. Since $3$ is a common direction for the NS5, $(1,k)5$ and D5-branes
transverse to the D3-branes, the left and right D3-brane stacks can slide freely and separately along $3$. 
Fig.\ \ref{D3D5}$a$ depicts the generic case of a stack of $N$ D3-branes on the left and $N+M$ D3-branes on
the right at different values of $x^3$. The 3-brane charge at the position of the defect, $x^2=0$, is absorbed by 
the $N+M$ D5-branes.

Let us consider first the low-energy dynamics in the open string sectors of Fig.\ \ref{D3D5}$a$. 
The massless modes of the blue 3-3 strings exclusively on the left or the right stack include 
two $\NN=1$ spinor multiplets, whose low-energy
dynamics is captured by $\NN=1$ CS theories at level $k$ and gauge group $U(N)$ and $U(N+M)$ 
respectively, and two $\NN=1$ scalar multiplets that capture the brane motion in direction 3. 
There are also 3-5 strings (denoted with a black color) that give rise to massless $\NN=1$ scalar multiplets 
denoted as $h$ and $g$. $h$, which originates from the left 3-5 strings, is in the bifundamental representation of $U(N)\times U(N+M)$. $g$ comes from the right 3-5 strings and is in the bi-fundamental of 
$U(N+M)\times  U(N+M)$.
Finally, there are red 3-3 strings stretching from the left D3-brane stack to the 
right D3-brane stack that give rise to $\NN=1$ scalar multiplets denoted as $P$. These fields are in 
the bifundamental of $U(N)\times U(N+M)$.
When the left and right D3-brane stacks are in different $x^3$ positions (as in Fig.\ \ref{D3D5}$a$) 
the scalar multiplet $P$ arises from a long string sector and is obviously massive. After $P$ is integrated out
one is left with a boundary theory $\tt L$ localized at the intersection with the $N$ D3-branes, and another
boundary theory $\tt R$ at the intersection with the $N+M$ D3-branes. $\tt L$ is a theory for the 
$U(N)\times U(N+M)$ bifundamental fields $h$. We propose that the bosonic part of this theory is given
by the action \eqref{generalab} with $\GG=U(N)$, $\GG'= U(N+M)$. The supersymmetric completion can 
be performed easily along the lines of section \ref{N1}. Similarly, $\tt R$ is a theory for the $U(N+M)\times U(N+M)$ 
bifundamental fields $g$. It can be formulated similarly along the lines of subsection \ref{general}.

%
A more interesting situation arises when the left and right D3-brane stacks are placed at the same $x^3$ position
as described in Fig.\ \ref{D3D5}$b$. In this case the $U(N)\times U(N+M)$ bifundamentals $P$ are massless
and it becomes possible to impose a different set of boundary conditions ---the boundary conditions 
\eqref{bosonca}-\eqref{bosoncaa} that `reconnect' the $N$ D3-branes on the left with $N$ D3-branes
on the right. This reconnection breaks the $U(N)_L \times U(N)_R$ symmetry subgroup to the diagonal 
$U(N)$. Accordingly, we propose that the low-energy theory of this configuration is described by the action 
\eqref{N1ba}, \eqref{N1bc} that gives rise to the $\NN=(1,0)$ $U(N+M)_k/U(N)_k$ gauged WZW model.

This type of reconnection and the proposed domain wall theory
can also be understood in the following manner. In the absence of the bifundamental fields $P$ and their 
interactions the total boundary theory is the direct sum of the left and right theories ($\tt L$, $\tt R$) in the 
situation of Fig.\ \ref{D3D5}$a$.
When the fields $P$ are massless a new 
branch of marginal deformations of the open string theory opens up where the bottom components of the fields 
$P$ acquire a vacuum expectation value (vev). When the vev, as an $N\times (N+M)$ matrix, is
\beq
\label{IR1aa}
\langle P \rangle = \mu \left(  {\boldsymbol 1}_{N\times N} ~~ {\boldsymbol 0}_{N\times M} \right)
\eeq 
the symmetry subgroup $U(N)_L\times U(N)_R$ reduces to the diagonal. 
At the same time, part of the $h$ fields from 
the left 3-5 sector become massive and can be integrated out. This is a consequence of 
a cubic boundary interaction
\beq
\label{IR1ab}
\int d\theta_-~ \Tr \left( g^{-1} \hat{\bar P} \, h + \bar h\, \hat P\, g \right)
~,
\eeq
where $\bar P$ is the Hermitian conjugate of the matrix $P$, and the hat denotes the boundary projection. 
When $P$ acquires the vev \eqref{IR1aa} only the $h$ modes of the form 
\beq
\label{IR1aba}
h_{\hat a}^{~\hat b} = \left[ \Pi_{U(N)} (g)\right]_{\hat a}^{~\hat b} ~, ~~
h_{\hat a}^{~ b_\perp} = 0
\eeq
do not receive a massive interaction. $\hat a$ are indices in the fundamental of $U(N)$ and $a_\perp$
are indices in the fundamental of $U(N+M)$ orthogonal to the $U(N)$ subgroup. Then, we can effectively
set $h$ in the profile \eqref{IR1aba} in the rest of the action and recover 
\eqref{N1ba}, \eqref{N1bc} for the 
$U(N+M)$ group superfields $g$ with 3d bulk spinor superfields $\Gamma_L$ in $U(N)$ and $\Gamma_R$ 
in $U(N+M)$.

\subsection{Bulk Seiberg duality and level-rank duality on the defect}
\label{branemotion}

Further evidence in favor of the above proposal can be obtained by analyzing the effects of other deformations
of the brane setup \eqref{setupaa}. We will focus on a deformation that
slides the $(1,k)$5-brane along the 
6-direction through the NS5-brane. This operation is commonly performed in HW setups to argue in favor
of Seiberg duality for the low-energy gauge theory on the brane configuration. Characteristic applications 
of this deformation include \cite{Elitzur:1997fh} in the context of type IIA HW setups and 4d $\NN=1$ super-QCD
theory, or \cite{Giveon:2008zn} in the context of type IIB HW setups and 3d $\NN=2$ Chern-Simons-SQCD theory.
In Ref.\ \cite{Armoni:2009vv} the exchange of the $(1,k)$5-brane with the NS5-brane along the 6-direction in the
brane configuration \eqref{setupaa} without any D5-branes was used to exhibit the IR equivalence 
(Seiberg duality) between the 3d $\NN=1$ CS theory at level $k$ with gauge group $U(N)$ 
---in short, $U(N)_k$ CS theory--- and its dual, the $U(|k|-N)_{-k}$ CS theory.\footnote{A similar brane exchange in a {\it non-SUSY} setup leads to a string theory realisation of level-rank duality between two purely bosonic Chern-Simons theories \cite{Armoni:2014cia}.}

In the presence of a domain wall, or boundary, we expect to see the bulk Seiberg duality to translate on the
two-dimensional defect to a corresponding duality symmetry. With a gauged WZW model on the defect the 
anticipated duality is some version of level-rank duality. Indeed, this expectation is borne out correctly by 
the above-proposed description of the brane setup \eqref{setupaa}.

With $N+M>0$ D5-branes the following effects are taking place. As we pass the $(1,k)$5-brane past the NS5-brane
$k$ D3-branes are created and the original $N$ and $N+M$ D3-branes are dragged along to become 
anti-D3-branes. The D5-branes remain unchanged. At this point there are $k$ D3 and $N$ anti-D3-branes
at $x^2<0$, and $k$ D3-branes with $N+M$ anti-D3-branes at $x^2>0$. 

After the annihilation of the corresponding D3-$\overline{\rm D3}$ pairs we obtain the brane 
configuration
\beq
\label{IR1ba}
	\begin{array}{r c c c c c c c c c c c}
	k-N-M ~ D3_+ & ~:~ & 0  & 1 & 2_+ &  &  &  & |6| &  &  & 
	\\
	k-N ~ D3_- & ~:~ & 0  & 1 & 2_- &  &  &  & |6| &  &  & 
	\\
	1 ~ NS5  & ~:~ & 0  & 1 & 2 & 3 & 4 & 5 &  &  &  & 
	\\
	1 ~(1,k)5 & ~:~ & 0  & 1 & 2 & 3 &  & \left[ {5 \atop 9} \right]_\theta &  &  & 8 & 
	\\
	N+M ~ D5  & ~:~ & 0  & 1 &  & 3 & 4 & 5 & 6 &  &  & 
	\end{array}
\eeq
Full annihilation to a supersymmetric configuration requires $k \geq N+M$. The resulting field theory has 
the same form as the original one with the substitution
\bea
& & k \rightarrow k \nonumber \\
& & M \rightarrow M \\
& & N \rightarrow k-N-M \label{duality-map}
\nonumber
\eea
and according to the proposal of the previous subsection \ref{IRtheory}
it gives rise at low energies to the 2d $\NN=(1,0)$ 
$$
\frac{U(k-N)_k}{U(k-N-M)_k}
$$ 
gauged WZW model.

With the common assumption that the IR physics of the brane configuration are insensitive to the above
brane operation we are led to the following equivalence between $\NN=(1,0)$ gauged WZW models on the domain
wall
\beq
\label{IR1bb}
\frac{U(N+M)_k}{U(N)_k} = \frac{U(k-N)_k}{U(k-N-M)_k}
~.
\eeq
Borrowing a standard language from Seiberg duality, we will call the LHS of this equivalence 
the `electric theory', and the RHS the `magnetic theory'.

This equivalence passes a number of immediate tests:
\begin{itemize}

\item[$(a)$] The global $U(M)$ symmetry matches on both sides.

\item[$(b)$] The central charges match non-trivially:
\bal
\label{IR1bc}
c_{electric} &= \left( \frac{3}{2} - \frac{N+M}{k} \right) \left( (N+M)^2-1 \right) 
- \left( \frac{3}{2} - \frac{N}{k} \right) \left( N^2 -1 \right)
\\
&= \left( \frac{1}{2}+\frac{N}{k} \right) \left( (k-N)^2 -1 \right) 
- \left( \frac{1}{2} + \frac{N+M}{k} \right) \left( (k-N-M)^2 -1 \right) = c_{magnetic}
~.\nonumber
\end{align}
We remind the reader that the central charge of the $U(N)_k$ supersymmetric WZW model is
\beq
\label{IR1bca}
c_{N,k}= \left( \frac{3}{2} - \frac{N}{k} \right) (N^2-1) + \frac{3}{2}
~.
\eeq

\item[$(c)$] In the special case with $N=0$ we obtain the duality relation 
\beq
\label{IR1bd}
U(M)_k = \frac{U(k)_k}{U(k-M)_k}
\eeq
which is a well known level-rank duality relation following from the triviality of the coset
$$
\frac{U(k)_k}{U(M)_k \times U(k-M)_k}
~.
$$
Superficially, the generalized level-rank duality \eqref{IR1bb} follows from \eqref{IR1bd} by applying it  
separately on the numerator and denominator of the coset
\beq
\label{IR1be}
\frac{U(N+M)_k}{U(N)_k} = \frac{U(k)_k \times U(k-N)_k}{U(k-N-M)_k \times U(k)_k}
= \frac{U(k-N)_k}{U(k-N-M)_k}
~.
\eeq

\item[$(d)$] The brane configuration suggests that the Witten index of the model is
\beq 
I={k \choose N+M}{N+M \choose N} \label{witten-index}
\eeq
The above formula is a consequence of the s-rule. It corresponds to the various ways the $N+M$ D3 branes can end on the $(1,k)$ fivebrane, times the various choices of selecting $N$ D3 branes inside the $N+M$ collection.

Note that the Witten index \eqref{witten-index} passes several non-trivial checks: $(i)$ it is invariant under the 
duality map \eqref{duality-map}, $(ii)$ it yields the correct Witten index for the $U(M)_k$ model when $N=0$, and 
$(iii)$ it yields the correct Witten index for the $U(M)_k$ model when $k=N+M$.
  
\end{itemize}

\section{1/16-BPS brane setups}
\label{N2brane}

The low-energy description of the 1/16-BPS configurations in subsections \ref{20setup}, \ref{11setup} involves 
domain walls in 3d $\NN=2$ CS theories. These can be treated as above by
employing the formalism of section \ref{N2}. Since many of the technical details are closely related with those
in section \ref{N2} we will be mostly brief in this section highlighting the novel features of $\NN=2$ supersymmetry.

An important difference between the $\NN=1$ configuration \eqref{setupaa} and the $\NN=2$ configurations 
\eqref{setupab}, \eqref{setupac} is that contrary to \eqref{setupaa} in \eqref{setupab} and \eqref{setupac} the 
suspended D3-branes are forced to lie at a specific position in the direction 3, say $x^3=0$. Accordingly, in the low 
energy description of the D3-brane theory there are no massless scalars capturing free motion in any of the 
directions of the $N+M$ D5-branes and the only possible configuration is the one depicted in Fig.\ \ref{D3D5}$b$.

For each of the configurations \eqref{setupab} or \eqref{setupac}, the low-energy bulk 3d gauge theory 
is $\NN=2$ CS theory at level $k$ with gauge group $U(N)$ on the left and $U(N+M)$ on the right. There 
are again bi-fundamental fields $P$, $h$ and $g$ as in section \ref{IRtheory}, which are now $\NN=2$ chiral
superfields. The interactions of these fields with bulk $\NN=2$ vector superfields and among themselves are
formulated following the rules of section \ref{N2} according
to the specifics of the supersymmetry preserved by the domain wall, that is 2d $\NN=(2,0)$ supersymmetry
in the brane configuration \eqref{setupab}, and 2d $\NN=(1,1)$ supersymmetry in the brane configuration 
\eqref{setupac}. With a non-zero vacuum expectation value \eqref{IR1aa} for the $\NN=2$ chiral superfield $P$ and 
an $\NN=2$ superpotential interaction of the form \eqref{IR1ab} the chiral superfields $h$ can be integrated
out to obtain the $\NN=(2,0)$ (in the case of \eqref{setupab}), and $\NN=(1,1)$ (in the case of \eqref{setupac})
$$
\frac{U(N+M)_k}{U(N)_k}
$$
gauged WZW model on the domain wall.

Repeating the brane argument for Seiberg duality of subsection \ref{branemotion} we recover
verbatim the $\NN=(2,0)$ or $\NN=(1,1)$ version of the generalized level-rank duality \eqref{IR1bb}.

\section{Outlook}
\label{outlook}

We described in field theory a general class of domain wall theories in 
(supersymmetric) CS theories. Some of these theories realize gauged WZW models. 
We demostrated that these constructions
appear naturally in the context of type IIB HW setups, where the two-dimensional domain walls arise by suitably
including D5-branes on which D3-branes end. Standard arguments of brane exchange
that realize 3d Seiberg dualities translate in this context to a corresponding statement of generalized 
level-rank duality on the domain wall theories.

The results presented in this paper pave the way towards a more general study of domain wall theories in 
Chern-Simons-matter theories and related type IIB brane constructions of these theories. The incorporation 
of matter to the story outlined in the previous sections can be performed in a rather straightforward manner.
A specific application to the ABJM model, the M2-M5 brane theory and its type IIB construction will be 
discussed in a companion paper. 

A class of examples that are interesting to analyze further along the lines of the present
work are domain wall theories in 3d $\NN=2$ CS theories with unitary gauge groups coupled to 
a number of fundamental and anti-fundamental chiral superfields. These theories can be engineered simply
in the HW setups \eqref{setupab}, \eqref{setupac} with the addition of an extra set of (flavor) D5-branes in the 
directions $(012789)$. When the 5-branes are exchanged in this setup the 3d bulk Chern-Simons-matter
theories undergo a Giveon-Kutasov duality \cite{Giveon:2008zn}. 
The domain wall theories in our setup will exhibit a corresponding
level-rank duality in the presence of additional matter fields and bulk-boundary couplings that is interesting
to analyze further.

Another interesting, potentially related, aspect of our work appears in the context of the brane setups 
\eqref{setupaa}. In the absence of the D5-branes that produce domain walls it was pointed out in 
\cite{Armoni:2009vv} that the low-energy 3d $\NN=1$ gauge theory is very closely related to the Acharya-Vafa 
(AV) theory \cite{Acharya:2001dz} that has been argued to describe the low-energy theory on the domain walls of 
the four-dimensional $\NN=1$ super-Yang-Mills (SYM) theory. Ref.\ \cite{Armoni:2009vv} explained how several 
known facts about the domain walls of the 4d $\NN=1$ SYM theory are reproduced correctly by the theory
on the suspended D3-branes in the HW setup \eqref{setupaa}. More recently, Gaiotto conjectured 
\cite{Gaiotto:2013gwa} that the theory on a general junction of domain walls of the 4d $\NN=1$ 
SYM theory is given by an $\NN=(1,0)$ gauged WZW coset. It would be interesting to see if such junctions 
can be realized in a natural way in the context of the HW setup \eqref{setupaa}, and if the conjectured 
gauged WZW coset of \cite{Gaiotto:2013gwa} can be reproduced from a direct analysis of the type 
presented above in this paper. The simplest example of a domain wall junction analyzed in \cite{Gaiotto:2013gwa} 
is one where a 4d SYM domain wall with $U(N)$ AV gauge theory (with CS level $k$) on the left meets on a 
two-dimensional defect a conjugate (or Seiberg dual) domain wall 
with $U(k-N)$ AV gauge theory that runs to the right. In this case the two-dimensional defect is a topological
Seiberg-duality wall for the AV theory, and \cite{Gaiotto:2013gwa} conjectured that it involves the $\NN=(1,0)$
$$
\frac{U(k)_k}{U(N)_k \times U(k-N)_k}
$$
gauged WZW coset. It would be interesting to obtain an explicit construction of this Seiberg-duality wall
theory.

\subsection*{Acknowledgments}

We would like to thank D.\ C.\ Thompson for useful discussions and a critical reading of the manuscript.
A.A.\ is grateful to the U.K.~Science and Technology Facilities Council
(STFC) for financial support under grants ST/J000043/1 and ST/L000369/1.
The work of V.N.\ was supported
in part by European Union's Seventh Framework Programme under grant agreements (FP7-
REGPOT-2012-2013-1) no 316165, the EU-Greece program ``Thales'' MIS 375734 and was
also co-financed by the European Union (European Social Fund, ESF) and Greek national
funds through the Operational Program ``Education and Lifelong Learning'' of the National
Strategic Reference Framework (NSRF) under ``Funding of proposals that have received a
positive evaluation in the 3rd and 4th Call of ERC Grant Schemes''.

\begin{appendix}

\section{Brane supersymmetries}
\label{braneSUSY}

The supersymmetries preserved by brane configurations of the type \eqref{setupaa}-\eqref{setupac} without the D5-branes have been studied extensively in the past \cite{Aharony:1997ju,Kitao:1998mf,Bergman:1999na}. 
Here we summarize the less studied effects of the $N+M$ D5-branes, that play the role of the domain wall in the 
low-energy gauge theory.

In ten-dimensional type IIB string theory the supersymmetry transformations are implemented by two 
real 16-component spinors of the same chirality, $\varepsilon_1,\varepsilon_2$. 
Each of the brane stacks in \eqref{setupaa}-\eqref{setupac}
project these spinors appropriately. Omitting standard details we summarize the resulting projection equations
in each case. $\Gamma_\mu$ $(\mu=0,1,\ldots,9)$ are 10d flat spacetime $\Gamma$-matrices. The notation
$\Gamma_{\mu_1\ldots\mu_k}$ refers to the product $\Gamma_{\mu_1}\cdots \Gamma_{\mu_k}$, which is
totally antisymmetric in its indices.

\paragraph{Brane setup \eqref{setupaa}.} We obtain the projection equations
\bal
\label{SUSYaa}
{\rm D3}~&: ~~ \Gamma_{0126}\, \varepsilon_1 = - \varepsilon_2 ~,
\\
\label{SUSYab}
{\rm NS5}~ &: ~~ \Gamma_{012345}\, \varepsilon_1= \varepsilon_1~, ~~
\Gamma_{012345}\, \varepsilon_2=- \varepsilon_2~, 
\\
\label{SUSYac}
{(1,k)5}~ &: ~~ \left(1-\cos\theta\, \gamma_{(6)} \right) \varepsilon_1 = \sin\theta\, \gamma_{(6)}\, \varepsilon_2 
~, ~~ \gamma_{(6)} \equiv \Gamma_{01238} \left( \cos\theta\, \Gamma_5+ \sin\theta\, \Gamma_9 \right)~,
\\
\label{SUSYad}
{\rm D5}~&:~~ \Gamma_{013456}\, \varepsilon_1 =  \varepsilon_2
~.
\end{align}
The angle $\theta$ obeys the relation
\beq
\label{SUSYada}
\tan\theta = k\, g_s
~.
\eeq

It is easy to check that the combined D3, NS5, $(1,k)$5 equations preserve only 2 supersymmetries.
With the use of these equations the last one coming from the D5-branes becomes
\beq
\label{SUSYae}
\Gamma_{01} \, \varepsilon_1 = - \varepsilon_1 
\eeq
reducing the supersymmetry by a further 1/2. This results to a 2d chiral spinor.

\paragraph{Brane setup \eqref{setupab}.}
The D3, NS5, D5 projection equations are the same as in \eqref{SUSYaa}, \eqref{SUSYab},
and \eqref{SUSYad}. The remaining $(1,k)5$ equation is
\bal
\label{SUSYafa}
(1,k)5~ : ~~ \left( 1-\cos\theta\, \gamma_{(6)} \right) \varepsilon_1 = \sin\theta\, \gamma_{(6)} \varepsilon_2 
~, ~~ \gamma_{(6)} \equiv \Gamma_{01289} \left( \cos\theta\, \Gamma_3 - \sin\theta\, \Gamma_7 \right)
~.
\end{align}
Without the D5-branes the configuration preserves four supersymmetries. A little manipulation shows that
the 16-component spinor $\varepsilon_1$ is projected twice
\beq
\label{SUSYag}
\Gamma_{012345}\, \varepsilon_1 = \varepsilon_1 ~, ~~ \Gamma_{4567}\, \varepsilon_1 = \varepsilon_1
~.
\eeq

The D5-branes reduce supersymmetry by a further 1/2 giving again the projection equation \eqref{SUSYae}.
We obtain $d=2$ $\NN=(2,0)$ supersymmetry.

\paragraph{Brane setup \eqref{setupac}.}
The only difference compared to the previous configuration is the D5 orientation that results to the spinor
projection equation
\beq
\label{SUSYba}
{\rm D5}~:~~ \Gamma_{016789}\, \varepsilon_1 = \varepsilon_2
~.
\eeq
There are no contraints on $\Gamma_{01}\, \varepsilon_1$ now, hence we obtain $\NN=(1,1)$ supersymmetry
in two dimensions.

\section{Supersymmetry conventions}
\label{SUSYconventions}

For quick reference and the convenience of the reader 
in this appendix we summarize our conventions for $\NN=1$ and $\NN=2$ supersymmetry in three dimensions.

\subsection{$\NN=1$ supersymmetry}

\subsubsection*{Spinor index manipulations}
We use small letters from the beginning of the Greek alphabet, $\alpha,\beta,\ldots$, to denote spinor indices. 
Small Greek letters from the middle of the Greek alphabet, $\mu,\nu,\dots$, are reserved for the 3d spacetime
indices. 

Spinor index manipulations of Grassmann odd variables $\theta_\alpha$ and products are performed according to 
the following rules
\bea
\label{1conaa}
\theta^\alpha &=& C^{\alpha \beta} \theta_\beta
~,
\\
\theta_\beta &=& \theta^\alpha C_{\alpha \beta}
~,
\\
\theta_\alpha \theta_\beta &=& - C_{\alpha \beta} \theta^2~, ~~ 
\theta^2 = \frac{1}{2}\theta^\gamma \theta_\gamma = \frac{1}{2} C^{\gamma \beta} \theta_\beta \theta_\gamma
~,
\\
C_{\alpha\beta} &=& -C_{\beta \alpha} = - C^{\alpha \beta} = \left( {0 \atop i } ~~{-i \atop 0} \right)
~,
\\
C_{\alpha\beta}C^{\gamma \delta} &=& \delta^\gamma_{[\alpha} \, \delta^{\delta}_{\beta]}
~,
\\
(\psi_1 \psi_2) &\equiv& \psi_1^\alpha \psi_{2\alpha}
~,
\\
\left( \gamma^\mu \theta\right)_\alpha &=& (\gamma^\mu)_\alpha^{~~\beta}\theta_\beta
~.
\eea
The $\gamma^\mu$ matrices obey the Clifford algebra
\bea
\label{1conab}
\{ \gamma^\mu,\gamma^\nu \}_{\alpha\beta} = 2\, \eta^{\mu\nu} C_{\alpha\beta}
~,~~
\gamma^\mu_{\alpha\beta} = (\gamma^\mu)_\alpha^{~\gamma} C_{\gamma\beta} = \gamma^\mu_{\beta\alpha}
~,
\eea 
where $\eta_{\mu\nu}$ is the 3d Minkowski metric.

\subsubsection*{SUSY algebra, covariant derivatives and superfields}
The supercharges $Q_\alpha$ of $\NN=1$ supersymmetry and the corresponding superspace derivatives 
$D_\alpha$ obey
\bea
\label{1conac}
\d_\alpha \theta_\beta &=& C_{\alpha \beta}~, ~~ \d_\alpha \theta^\beta =\delta_\alpha^\beta
~,\\
Q_\alpha &=& \d_\alpha - \left( \gamma^\mu \theta \right)_\alpha \d_\mu
~,\\
D_\alpha &=& \d_\alpha + (\gamma^\mu \theta)_\alpha \d_\mu
~,\\
\{Q_\alpha, Q_\beta\} &=& - \{ D_\alpha,D_\beta \} = 2\gamma_{\alpha \beta}^\mu \d_\mu
~.
\eea

The two basic multiplets of $\NN=1$ supersymmetry can be formulated with the use of the off-shell superfields
\bea
\label{1conad}
{\rm Scalar}~~&:&~~ \Phi = \phi + (\theta \psi) + \theta^2 F
~,
\\
{\rm Spinor}~~&:&~~ \Gamma_\alpha = \chi_\alpha + \theta_\alpha M 
+(\gamma^\mu \theta)_\alpha A_\mu + \theta^2 \left( \lambda_\alpha - (\gamma^\mu \d_\mu \chi)_\alpha \right)
\eea
The spinor superfields, in particular, are used to package the gaugino $\lambda_\alpha$ and the gauge field 
$A_\mu$. The Majorana spinor $\chi_\alpha$ and the real scalar $M$ are auxiliary fields. In the formulation
of the $\NN=1$ CS theory \eqref{N1ab} it is also useful to define the related superfields
\bea
\label{1conada}
\Omega_\alpha &=& \omega_\alpha - \frac{1}{6} \left[ \Gamma^\beta, \Gamma_{\alpha\beta} \right]
~, 
\\
\omega_\alpha &=& \frac{1}{2} D^\beta D_\alpha \Gamma_\beta 
+\frac{1}{2} \left[ \Gamma^\beta, D_\beta \Gamma_\alpha \right] 
+\frac{1}{6} \left[ \Gamma^\beta, \{ \Gamma_\beta, \Gamma_\alpha \} \right]
~,
\\
\Gamma_{\alpha\beta} &=& \frac{1}{2} \left[ D_{(\alpha} \Gamma_{\beta)} + \{ \Gamma_\alpha, \Gamma_\beta\} \right]
~.
\eea

In the case of gauge theories it is also convenient to form the super-gauge-covariant derivative
\bea
\label{1conae}
\nabla_\alpha = D_\alpha + \Gamma_\alpha
~.
\eea
The conventional gauge-covariant derivative, which is the $(\gamma^\mu \theta)_\alpha$ 
component of $\nabla_\alpha$, is denoted as
\bea
\label{1conaf}
\DD_\mu = \d_\mu + A_\mu
\eea
in the main text.

\subsection{$\NN=2$ supersymmetry}
\label{2conventions}

For $\NN=2$ supersymmetry we sometimes use the complex Grassmann variables
\beq
\label{2conaa}
\vartheta_\alpha = \frac{1}{\sqrt 2}(\theta_{1\alpha}+i \theta_{2\alpha})
~.
\eeq
For Grassmann integrations
\beq
\label{2conab}
\int d^4\vartheta = \int d^2\vartheta d^2 \vartheta = - \int d^2 \theta_1 d^2 \theta_2
~.
\eeq
Accordingly, we define the $\NN=2$ superspace covariant derivatives as
\beq
\label{2conaba}
{\boldsymbol D}_\alpha = \d_\alpha + (\gamma^\mu \bar \vartheta)_\alpha \d_\mu ~, ~~
\bar {\boldsymbol D}_\alpha = \bar\d_\alpha + (\gamma^\mu \vartheta)_\alpha \d_\mu
\eeq
which can be decomposed to the $\NN=1$ superspace derivatives
\beq
\label{2conabb}
{\boldsymbol D}_\alpha = \frac{1}{\sqrt 2} \left( D_1 -i D_2\right) ~, ~~
\bar {\boldsymbol D}_\alpha =\frac{1}{\sqrt 2} \left( D_1 + i D_2 \right)
~.
\eeq

\subsubsection*{Superfields}

The main superfields of $\NN=2$ supersymmetry are vector superfields and chiral superfields.

An $\NN=2$ vector superfield $V$ can be built out of two $\NN=1$ scalar superfields $A,B$ and
an $\NN=1$ spinor superfield $\Gamma_\alpha$ as
\beq
\label{2conac}
V(\theta_1,\theta_2) =  A(\theta_1) + \theta_2^\alpha \Gamma_\alpha (\theta_1)
+\theta_2^2 \big( B(\theta_1) + D_1^2 A (\theta_1) \big)
~.
\eeq
In the so-called Ivanov gauge one sets $A=0$ and uses the Wess-Zumino gauge $(\chi = M=0)$ for 
$\Gamma_\alpha$.

An $\NN=2$ chiral superfield $\Phi$ can be built out of a complex $\NN=2$ scalar superfield $X$ as
\beq
\label{2conad}
\Phi (\theta_1,\theta_2) = X(\theta_1) + i \theta_2^\alpha D_{1\alpha} X + \theta_2^2  D_1^2 X
~.
\eeq

In the presence of a gauge symmetry the chiral $\NN=2$ super-gauge-covariant derivative takes the form
\beq
\label{2conae}
{\boldsymbol \nabla}_\alpha = e^{-V} {\boldsymbol D}_\alpha e^V
~.
\eeq

\end{appendix}

\newpage
\providecommand{\href}[2]{#2}\begingroup\raggedright

\end{document}